\documentclass[conference,a4paper]{IEEEtran}

\usepackage[utf8]{inputenc} 
\usepackage[T1]{fontenc}
\usepackage{url}              
\usepackage{cite}             

\usepackage[cmex10]{amsmath}  
\usepackage{amssymb}
\usepackage{mleftright}       
\mleftright                   

\usepackage{graphicx}         
\usepackage{booktabs}         
\usepackage{comment}
\ifCLASSOPTIONcompsoc 
\usepackage[caption=false,font=normalsize,labelfont=sf,textfont=sf]{subfig}
\else
\usepackage[caption=false,font=footnotesize]{subfig}
\fi
\usepackage[linesnumbered,boxed, vlined]{algorithm2e}  

\usepackage[bookmarks=false]{hyperref}
\usepackage{xcolor}

\begin{document}
	
	\title{RECIPE: Rateless Erasure Codes Induced by Protocol-Based Encoding} 
	
		\author{\IEEEauthorblockN{Jingfan Meng}
				\IEEEauthorblockA{\textit{Georgia Tech} \\
						jfmeng@gatech.edu}
				\and
				\IEEEauthorblockN{Ziheng Liu}
				\IEEEauthorblockA{\textit{University of Utah} \\
						ziheng.liu@utah.edu}
				\and
				\IEEEauthorblockN{Yiwei Wang}
				\IEEEauthorblockA{\textit{Georgia Tech} \\
						ywang3607@gatech.edu }
				\and
				\IEEEauthorblockN{Jun Xu}
				\IEEEauthorblockA{\textit{Georgia Tech} \\
						jx@cc.gatech.edu}
			}

\newtheorem{theorem}{Theorem}[section]
\newtheorem{lemma}[theorem]{Lemma}
\newtheorem{proposition}[theorem]{Proposition}
\newtheorem{corollary}{Corollary}
\newtheorem{definition}[theorem]{Definition}
\newtheorem{conjecture}{Conjecture}[section]
\newtheorem{example}[theorem]{Example}
\newtheorem{remark}{Remark}

\newcommand{\ABBR}{RECIPE}
\newcommand{\Name}{Rateless Erasure Codes Induced by Protocol-based Encoding}
\renewcommand{\sectionautorefname}{Section}
\renewcommand{\subsectionautorefname}{Section}
\renewcommand{\algorithmautorefname}{Alg.}
\renewcommand{\equationautorefname}{Eq.}
\newcommand{\definitionautorefname}{Def.}
\RestyleAlgo{ruled}
\SetKwComment{Comment}{/* }{ */}
\newcommand{\padd}[2]{p_A(#1,#2)} 
\newcommand{\pskip}[2]{p_S(#1,#2)} 
\newcommand{\preplace}[2]{p_R(#1,#2)} 
\newcommand{\Add}{\texttt{Add}} 
\newcommand{\Skip}{\texttt{Skip}} 
\newcommand{\Replace}{\texttt{Replace}} 

\SetEndCharOfAlgoLine{;}
	\maketitle
	
%
	\begin{NoHyper}
\begin{abstract}
LT (Luby transform) codes are a celebrated family of rateless erasure codes (RECs). 
Most of existing LT codes were designed for applications in which a centralized encoder possesses all message blocks and is solely responsible for encoding them into codewords.
Distributed LT codes, in which message blocks are physically scattered across multiple different locations (encoders) that need to collaboratively perform the encoding, has never been systemically studied before despite its growing importance in applications.
In this work, we present the first systemic study of LT codes in the distributed setting, and make the following three major contributions.
First, we show that only a proper subset of LT codes are feasible in the distributed setting, and give the sufficient and necessary condition for such feasibility.
Second, we propose a distributed encoding protocol that can efficiently implement any feasible code. The protocol is parameterized by a so-called action probability array (APA) that is only a few KBs in size, and any feasible code corresponds to a valid APA setting and vice versa.
Third, we propose two heuristic search algorithms that have led to the discovery of feasible codes that are much more efficient than the state of the art.  

\end{abstract}

\section{Introduction}
\label{sec:intro}

Rateless erasure codes (REC)~\cite{byers_digital_1998} are a powerful tool for reliable data transmission.
LT (Luby transform) codes~\cite{luby_lt_2002} are the best-known family of REC.  
LT codes are attractive for network applications, because they have both high coding efficiency and low decoding time complexity (using the peeling algorithm~\cite{jiang_parallel_2017}).
Most existing LT codes were designed for applications in which a 
centralized encoder possesses all  message blocks and is solely responsible for encoding them into codewords.
However, the past two decades has seen some applications 
in which the message blocks are physically scattered across multiple different locations (encoders) that need to collaboratively perform the encoding.


In this work, we perform the first systemic study of LT codes in the distributed setting, by posing three research questions.
Our starting question is ``Does the distributed setting impose certain additional constraints that make some LT codes not realizable (feasible)?''  
Since the answer to this question is yes, as we will explain shortly, it leads to two more questions.  
Our second question is ``Can we design an distributed encoding protocol that can be parameterized to realize (implement) any feasible LT code, and has 
low computational and storage overheads for the encoders?'' 
Our third question is ``Can we find good codes (with high coding efficiency) within this restricted (feasible) family?''




In this work, we make three major contributions by definitively answering these three questions, respectively.
Our first contribution is to show that only a proper subset of LT codes are feasible in the distributed setting, and to give the sufficient and necessary condition for an LT code to be feasible. Our second contribution is to propose a distributed encoding protocol that can efficiently implement any feasible code.
Hence, we call both this protocol, and the family of codes it generates, RECIPE (Rateless Erasure Codes Induced by Protocol-based Encoding). 
We say a code is RECIPE-feasible if it belongs to this family.  The ultimate goal of the RECIPE coding theory is to discover RECIPE-feasible codes that can achieve high coding efficiencies.  
However, to search for such codes appears challenging, as we will elaborate in \autoref{sec:parameter}.
Our third contribution is two heuristic search algorithms\footnote{We will release all source codes and resulting XDD distributions on \url{https://cc.gatech.edu/home/jx}.} that have led to the discovery of RECIPE-feasible codes that are much more efficient than the state of the art~\cite{ben_basat_pint_2020}.

\section{Background and Formulation}
\label{sec:background}

\subsection{Centralized LT codes and \textit{XOR} Degree Distributions}
\label{subsec:lt_code}

In this section, we provide a brief introduction to centralized LT coding concepts, terms, and notations.
Let $U$ be the set of message blocks to be encoded (for transmission) and $k = |U|$;  $k$ is called block size in the literature.
Each LT codeword is an \textit{XOR} sum of $d$ distinct, randomly selected, message blocks.
We refer to this set (of message blocks) as the XOR-set  in the sequel.
This random selection is required to be uniform in the sense that, for any $i\le k$, all ${k\choose i}$ ways (of selecting $i$ distinct message blocks from $U$ to XOR together) must be equally likely.
We refer to this requirement as the uniformity condition in the sequel. 
This $d$, which  is called the \textit{\textit{XOR} degree}, is in general a random variable.  
We call the probability distribution of $d$ XOR \textit{degree distribution} (XDD) and denote it as $\vec{\mu}$ in this paper.
Thanks to the uniformity requirement, any LT coding scheme is uniquely determined and hence defined by its XDD $\vec{\mu}$.  
Throughout this paper, we write a rightward arrow on the top of $\vec{\mu}$ to emphasize that it is a vector of $k$ scalars, in which the $i^{th}$ scalar is denoted by $\mu (i)$.

\subsection{Path Tracing:  Our Distributed LT Coding Problem}
\label{subsec:pint}


In this section, we describe the path tracing problem that our RECIPE coding theory is designed to tackle.  
Probabilistic in-band network telemetry (PINT) is an emerging protocol framework for real-time data center network (DCN) measurement and monitoring~\cite{kim_-band_2015, ben_basat_pint_2020}.
An important PINT task is path tracing~\cite{jeyakumar_millions_2014,handigol2014know,tammana2016simplifying}.
In path tracing, each participating switch (router) probabilistically encodes its (switch) ID 
into a dedicated ``PINT field'' contained in each packet
transiting through the switch;  this field is typically short (say no more than 16 bits) to keep the bandwidth overhead of the path tracing operation small.  
The goal of path tracing is for the destination host of a flow of packets to recover the entire network path this flow had traversed,  
from the codewords (``PINT field'' values) of these packets. 

Formally, path tracing can be modeled as a distributed LT coding problem as follows.
Consider a path tracing instance in which a source node SRC sends a flow of packets
to a destination node DST, following a path that contains switches $1$, $2$, $\cdots$, $k$ in that order.  
This instance corresponds to a coding instance, 
in which the $k$ switches are the encoders that each possesses a message block (which is its own switch ID), and DST 
is the decoder.  The ``PINT field'' in each packet is a codeword-in-progress when the packet traverses along the path from SRC to DST.
The distributed coding problem is how these $k$ switches should collaboratively encode each such codeword so that  DST
can recover the entire path from as few packets (codewords therein) as possible (i.e., achieve high coding efficiency), which is important for such an LT code to be useful 
in a data center network environment in which the vast majority of flows contain no more than several packets.

In this instance, each switch $i$ along the path can ``do something'' to a codeword-in-progress $C$ only during the packet's ``brief stay'' at $i$.
In other words, whatever $i$ decides to do to $C$, it is a one-shot online decision.
As shown in~\cite{ben_basat_pint_2020}, there are only three conceivable LT coding actions $i$ can do to $C$ (probabilistically):  \Add{} (XOR its ID $M_i$ with $C$), \Skip{} (do nothing to $C$), and \Replace{} ($C$ with $M_i$).  
These three actions are however sufficiently expressive, since it will become intuitive to readers that 
they themselves do not constrain in any way how ``large and rich'' the RECIPE-feasible family (of codes) can be.

Rather, what makes this coding problem challenging and ``shapes'' the RECIPE-feasible family is the following constraint
imposed by the data center network environment.  The constraint is that every switch (encoder) has to be stateless and ``weightless'' in the 
sense it performs the same extremely simple (``weightless'') LT encoding processing
on every transiting packet without \textit{consciously knowing} which path (coding instance) 
this packet travels (belongs to) or the path length.   This constraint is necessary 
for a path tracing operation to incur minimal systems overheads 
at data center switches, since at any moment a switch may be on 
the paths of millions of different source-destination flows, each of which corresponds to a different coding instance. 
Under this constraint, every switch has to probabilistically perform \Add{}, \Skip{}, or \Replace{}, with the same probability parameter settings (that we will elaborate in \autoref{sec:scheme}), on
each packet (the codeword-in-progress therein) transiting through it, independently.


We now elaborate, using the coding instance above, on the aforementioned ``not knowing the path length'', since it is arguably the most consequential part of the stateless constraint.
In general, any switch $i$, $i = 1, 2, \cdots, k$, does not know the exact value of $k$ (the block size of this instance) when processing a packet (codeword-in-progress) belonging
to this instance.  To be more precise, switch $i$ knows its ``position'' $i$ (from the time-to-live (TTL) field in the IP header of the packet), but not $k - i$ (``how far away'' the packet is from DST).
This constraint implies that, at any switch $i$ ($1\le i\le k$) in this instance, the codewords-in-progress (contained in the packets of the flow) after being processed by switches $1$ through $i$ must be (the realizations of) a valid 
code in the sense the (XOR-set distribution of the) code satisfies the uniformity condition and hence can be characterized by an XDD that we denote as $\vec{\mu}_i$.  
This is because any switch $i$ could be the last switch in another coding instance $\Gamma$ (of length $i$), and in this case the codewords-in-progress that switch $i$ (in instance $\Gamma$) transits are the final codewords.
This constraint also implies that 
a (parameterized) RECIPE protocol should induce the same XDD $\vec{\mu}_i$ for all instances (in the network) of size (length) $i$, since the probabilistic ``decision logic'' (whether the action should be \Add{}, \Skip{}, or \Replace{}) at switches 1 through $i$ 
in all such instances are identical due to the stateless constraint.  
Hence a RECIPE protocol, which is run concurrently by numerous coding instances of different path lengths in a data center network,
generally induces a sequence of XDDs $\vec{\mu}_1, \vec{\mu}_2, \cdots, \vec{\mu}_{K}$ (one for each path length), where $K$ is the maximum path length (i.e., diameter) in the network.  
Our aforementioned first contribution is to discover the necessary and sufficient condition these $K$ XDDs need to satisfy for their 
``concatenation'' to become a RECIPE-feasible code.   

Throughout this work, for ease of presentation, we assume that the PINT field in each packet 
has the same length as the ID of a switch, and that an LT codeword is the XOR sum of switch IDs. 
The PINT paper~\cite{ben_basat_pint_2020}
has proposed several solutions for cases in which the PINT field is 
shorter, such as hash-compressing the switch IDs or fragmenting a switch ID across multiple packets.

\subsection{PINT: The State of the Art Path Tracing Code}\label{sec:PINT}

The only prior work on the topic of distributed LT coding for path tracing is PINT~\cite{ben_basat_pint_2020}, which proposed a reasonably good code, but did not develop any theory.
In comparison, this work gets to the bottom of the problem, explores the entire design space, and reaps the reward of discovering much more efficient codes than
the PINT code.

The PINT code can be considered a linear combination of two (what we now call) RECIPE codes (XDD sequences).
In the first code, each packet, when arriving at its destination, carries 
in its codeword a uniformly and randomly chosen switch ID along its path.
This code is produced by switches performing reservoir sampling.  
The second code is produced by every switch \textit{XOR}-ing its switch ID to the codeword contained in an arriving packet with a fixed probability $p$.
In the resulting code (XDD sequence), each $\vec{\mu}_k$, $k = 1, 2, \cdots, K$, is precisely $Binomial(k, p)$.  
Neither code is efficient.
PINT~\cite{ben_basat_pint_2020} uses a linear combination of these two inefficient codes that is, surprisingly, much more efficient than both.

\section{RECIPE and Its Variant}
\label{sec:scheme}



In this section, we first present (in \autoref{sec:recipe-d}) the baseline RECIPE protocol, called RECIPE-d, that requires each codeword-in-progress $C$ to be accompanied with the value of $d$, the current XOR-degree of $C$.  This requirement increases the coding overhead by a few (say 6)
bits per packet (as $d<64$ in any current communication network).  
This extra coding overhead can be eliminated using a streamlined variant of RECIPE that we call RECIPE-t and present in \autoref{subsec:simulation_table}.
RECIPE-t can induce any RECIPE-feasible code approximately but accurately, at the tiny cost of requiring each encoder (switch) to store a small (no more than 1MB in size) precomputed table.  
RECIPE-d and RECIPE-t are the second aforementioned contribution of this work.

\subsection{Degree-Based RECIPE (RECIPE-d)}
\label{sec:recipe-d}

\begin{algorithm}
	Retrieve hop count $i$, \textit{XOR} degree $d$, and codeword $C$ from $pkt$\;
	$\nu \gets h(i, pkt)$; \tcp*[f]{$\nu\in [0, 1)$}\label{line:random-nu}\\
	\uIf{$\nu < \padd{i}{d}$\label{line-cond-begin}}{$C \gets C \oplus M$;  \tcp*[f]{Add action}}
	\uElseIf{$\nu < \padd{i}{d} + \preplace{i}{d}$}{$C \gets M$; \tcp*[f]{Replace action}}
	\uElse{Do nothing; \tcp*[f]{Skip action}\label{line-cond-end}}
	Update $d$ accordingly\;
	Write $d$ and $C$ back to the packet.
	\caption{RECIPE-d protocol by switch $i$.}
	\label{alg:recipe_d}
\end{algorithm}

\autoref{alg:recipe_d} shows 
how a switch whose ID is $M$ processes a packet $pkt$ (the codeword-in-progress $C$ therein) transiting through it, using
the RECIPE-d protocol (that is run by all switches with the same parameter setting). 
As shown in Lines 3 through 9, the switch performs one of the three aforementioned actions (\Add{}, \Skip{}, and \Replace{}) on $C$ 
with probability $p_A(i, d)$, $p_S(i, d)$, and $p_R(i, d)$ respectively.  Here $i$ is how far (in number of hops) this switch is away from the SRC of $pkt$,
which as mentioned earlier can be inferred from $pkt$'s TTL;  $d$ is the current XOR-degree of $C$ (that RECIPE-d ``pays'' to know as mentioned earlier). 
As such, the RECIPE-d protocol is parameterized by the 2D array $(p_A(i, d), p_S(i, d), p_R(i, d))$, $i=0, 1, \cdots, K$ and $d=1,2, \cdots,i-1$, that we call 
the action probability array (APA).  How to set (probability values in) APA so that the resulting RECIPE-d protocol induces 
a valid and a good code will be studied in \autoref{sec:algo} and \autoref{sec:parameter}, respectively.   
RECIPE-d is stateless since the 
(random) action the switch performs on $pkt$ depends only on $i$ and $d$, but not on the flow (coding instance) $pkt$ belongs to.

In any LT coding scheme, to decode a set of codewords, the host (in our case the DST) must know the XOR-set of each codeword in the set.  
\autoref{alg:recipe_d} uses a standard (in computer science) derandomization technique called global hashing that allows the DST of $pkt$ to 
recover the XOR-set of $C$ as follows.   A global hash function $h(\cdot, \cdot)$ is shared among all switches and hosts in the network.
As shown in Lines 2 through 9, the exact realized action this switch performs on $pkt$ is determined by the hash value $\nu = h(i, pkt)$.  
When $pkt$ reaches DST, DST can infer the XOR-set of $C$ therein from the hash values $h(i, pkt)$, $i = 1, 2, \cdots$, that DST itself can 
compute.

\subsection{Table-based RECIPE (RECIPE-t)}
\label{subsec:simulation_table}

Consider a hypothetical path of maximum possible length $K$ and a packet $pkt$ that travels down the hypothetical path to its DST. 
We define the following (random) action vector $\overrightarrow{act} \triangleq (act(1), act(2), \cdots, act(K))$, where each $act(i)$, $i = 1, 2, \cdots, K$, is the random action that switch $i$ performs on $pkt$ (the codeword $C$ therein).
Recall that in RECIPE-d, switch $i$ realizes only $act(i)$ according to the hash value $\nu=h(i, pkt)$ and the APA that parameterizes the protocol.  
The idea of RECIPE-t is to let every switch store an (identical) copy of a precomputed (via Monte-Carlo simulation of RECIPE-d) action vector sample table (AVST) whose rows are independent realizations of the 
random vector $\overrightarrow{act}$.   Suppose the AVST has $L$ rows (independent samples) that we denote as $\overrightarrow{act}_1$, $\overrightarrow{act}_2$, $\cdots$, $\overrightarrow{act}_L$.
When $pkt$ travels down its path, all switches along the path collaboratively sample one uniformly random (across $[L] \triangleq \{1, 2, \cdots, L\}$) row $l$ in AVST (i.e., $\overrightarrow{act}_l$); and for $i = 1, 2, \cdots$, switch $i$ performs 
$act_l(i)$ on the codeword contained in the packet.  This sampling (of $l$) is done collaboratively using a different global hash function $g(pkt)$ (than $h(i, pkt)$).  
In theory, when $L$ tends to infinity, the (approximate) XDD sequence RECIPE-t induces converges to the actual XDD it tries to ``simulate''.  
In practice, $L = O(10^4)$ is large enough to achieve a very close approximation, as will be shown in~\autoref{fig:zoomin}.

\section{RECIPE-Feasible XDD Sequences}
\label{sec:algo}
In this section, we state the first aforementioned contribution of this paper:  the sufficient and necessary condition for an XDD sequence to be RECIPE-feasible.
The sufficiency proof also explains how APA should be set to induce a RECIPE-feasible XDD sequence.
The RECIPE coding theory in this and the next two sections will be developed for RECIPE-d, with the understanding that 
we can approximate any parameterized (by APA) RECIPE-d using its streamlined variant RECIPE-t.

To begin with, we introduce a notation that makes our presentation easier. 
Consider any XDD $\vec{\mu}_i$ in the XDD sequence.
Recall that the uniformity condition means that for any $d\le i$, 
the probability for every size-$d$ subset of $[i]$ to be the XOR-set of $C$ is identical.
We denote this probability by $q_i(d)$, which is equal to $\mu_i(d)/{i \choose d}$ by definition.
Throughput this section, we denote an XDD sequence by $\vec{q}_1, \vec{q}_2, \cdots, \vec{q}_K$ instead of $\vec{\mu}_1, \vec{\mu}_2, \cdots, \vec{\mu}_K$. 

The following theorem shows that the family of RECIPE-feasible codes corresponds to a  ($K(K+1)/2$)-dimensional polytope bounded by the following linear constraints (facets).


\begin{theorem}
	\label{main_result}
	An XDD sequence $\vec{q}_1, \vec{q}_2, \cdots, \vec{q}_{K}$ is RECIPE-feasible if and only if for any $2 \leq i \leq K, 1 \leq d \leq i-1$, it holds that
	\begin{align} 
			q_{i-1}(d) \geq q_i(d)+q_i(d+1).
			\label{eq:ineq}
	\end{align}
\end{theorem}

The necessity part of \autoref{main_result} is proved in Appendix~\ref{pf-th31}, and its sufficiency part follows from the following APA designation that instantiates any RECIPE-feasible XDD sequence $\vec{q}_1, \vec{q}_2, \cdots, \vec{q}_K$. 
\begin{itemize}
	\item For $i = 1$, the first switch always replaces the (initially empty) codeword by its ID, like in PINT.
	In other words, we let $\padd{1}{0}=0$, $\pskip{1}{0}=0$, and $\preplace{1}{0}=1$.
\item For any $2 \leq i \leq K$ and $1 \leq d \leq i-1$, we let
\begin{equation}
	\label{eq:op_prob}
	\begin{split}
		\padd{i}{d}&=q_i(d+1) / q_{i-1}(d), \\
		\pskip{i}{d}&=q_i(d) / q_{i-1}(d), \\
		\preplace{i}{d}&=1-\padd{i}{d}-\pskip{i}{d}.\\
	\end{split}
\end{equation}
\end{itemize}


In the interest of space, we prove in Appendix~\ref{pf-sufficiency} that the APA entries in (\ref{eq:op_prob}) are well-defined for any RECIPE-feasible XDD sequence, and that a RECIPE-d protocol thus parameterized satisfies the uniformity condition and 
induces the XDD sequence $\vec{q}_1,\vec{q}_2,  \cdots, \vec{q}_K$.

\section{Search for Good XDD Sequences}
\label{sec:parameter}



It is very challenging to discover efficient codes in the RECIPE-feasible family for two reasons.  First, the search for good codes has to 
work with the aforementioned ($K(K+1)/2$)-dimensional RECIPE-feasible polytope.  
Second, for a distributed LT code (XDD sequence) $\vec{\mu}_1, \vec{\mu}_2, \cdots, \vec{\mu}_{K}$
to be considered good (in terms of coding efficiency), every $\vec{\mu}_k$, $k = 1, 2, \cdots, K$, needs to be good,   
because
if $\vec{\mu}_{k^*}$ for a certain path length $k^*$ is bad, then all coding instances of path length $k^*$ have low efficiency.


In this section, we propose two heuristic algorithms for searching for good RECIPE-feasible codes.    
The first algorithm, called HRS and to be described in \autoref{subsec:backward}, searches over the entire RECIPE-feasible polytope.
The second, called QPS and to be described in \autoref{subsec:extend}, searches over a much smaller polytope called invariant (RECIPE-feasible) sequences, but 
is much more computationally efficient in exploring the smaller polytope.  As a result, when $K$ is large (say in hundreds), only QPS can output good codes (on every $k$) ``in due time''.
We will plot the XDD of a ``good'' RECIPE code found by QPS in Appendix~\ref{sec:more-evaluation}.

\subsection{Heuristic Reversed Search (HRS)}
\label{subsec:backward}


Our first search algorithm, called heuristic reversed search (HRS), is to greedily solve a series of $K$ subproblems that each has $O(K)$ variables to work with.
The idea is to find good XDD's hop-by-hop in the reversed order (from last to first) while conforming to (\ref{eq:ineq}).
We start with an XDD $\vec{\mu}_K$ at the last hop. 
The default choice is Robust Soliton~\cite{luby_lt_2002} due to its high coding efficiency.
Then, for $i=K, K-1, \cdots, 2$, we iteratively search for a good XDD $\vec{\mu}_{i-1}$ (for one hop earlier) in the RECIPE-feasible region  (that satisfies~(\ref{eq:ineq}) under current $i$ and XDD $\vec{\mu}_i$).  It is straightforward to show that 
every subproblem thus formulated is feasible. 

HRS is reasonably computationally efficient for $K$ values that are not too large (say $K\le 128$). 
However, since the search for $\vec{\mu}_{i-1}$ depends on $\vec{\mu}_i$, $\vec{\mu}_{i-1}$ ``inherits'' any coding inefficiency of $\vec{\mu}_i$. 
As a result, when $K$ is larger than 100 or so, $\vec{\mu}_k$ found by HRS are not very efficient except when $k$ gets close to (the last hop) $K$.
\subsection{Quadratic Programming Search (QPS)}
\label{subsec:extend}

Our second scheme, called quadratic programming search (QPS), searches over only invariant (XDD) sequences (defined next) in the RECIPE-feasible polytope.   
\begin{definition}
An XDD sequence is \emph{invariant} (at each hop) if and only if for every XOR degree $d=1, 2, \cdots, K-2$, $\mu_{d+1}(d) = \mu_{d+2}(d)  = \cdots = \mu_K(d)$. 
\end{definition}
By this definition, every invariant XDD sequence is fully parameterized by the $K$ scalars in $\vec{\mu}_K$ (thus we drop the subscript $K$ in the following theorem). 
Furthermore, as a direct corollary of \autoref{main_result}, the following theorem shows that 
the subspace of XDD sequences that are both \emph{invariant and RECIPE-feasible} is a $K$-dimensional polytope.
\begin{theorem}\label{th:inv-point}
	An invariant XDD sequence is RECIPE-feasible if and only if for every $d=1, 2, \cdots, K-2$, $\mu(d) \geq (d+1)/d \cdot \mu(d+1)$.
\end{theorem}

\begin{figure*}[!ht]
	\centering
	\includegraphics[width=0.8\linewidth]{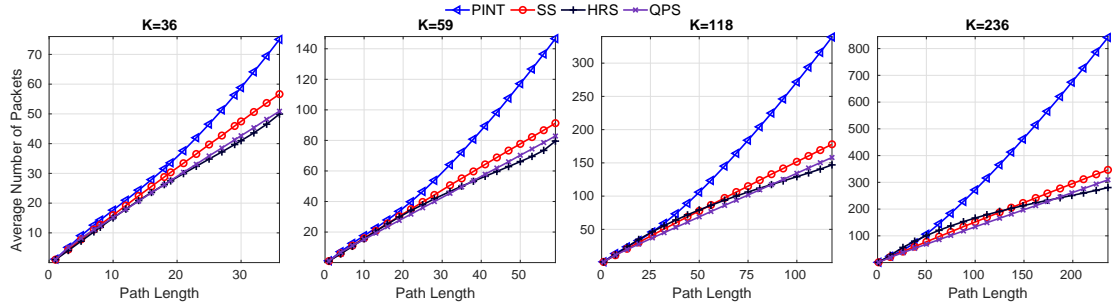}
	\caption{RECIPE codes vs PINT in terms of coding efficiency.}
	\label{fig:compare}
\end{figure*}
\begin{figure}[!ht]
	\centering\includegraphics[width=0.8\linewidth]{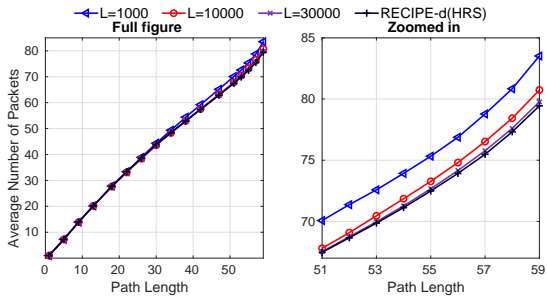}
	\caption{RECIPE-t compared against RECIPE-d.}
	\label{fig:zoomin}
\end{figure}

\begin{example}
	It is not hard to verify that the truncated (at $k = 1, 2, \cdots, K$) Soliton distributions (see~\cite{luby_lt_2002}), when concatenated into an XDD sequence, is neither 
	invariant nor RECIPE-feasible.  However, the following truncated XDDs $\vec{\mu}_k$, $k = 1, 2, \cdots, K$, when concatenated into an XDD sequence, is 
	both invariant and RECIPE-feasible. We call this XDD sequence \emph{Shifted Soliton}, since each $\vec{\mu}_k$ is a ``cyclic shift'' of the truncated (at $k$) Ideal Soliton.  
	We accidentally discovered Shifted Soliton, which in turn inspired us to propose QPS (to search for even better invariant sequences using computer).  
	\begin{align*}
		&\mu_k(d) = 1/ [d(d+1)],\: d = 1, 2, ..., k-1. \\
		&\mu_k(k) = 1/k.
	\end{align*}
\end{example}

Since QPS searches (using a quadratic programming procedure as we will describe in Appendix~\ref{expl-mean-field}, which gives QPS its name) over a $K$-dimensional ``sub-polytope'' (of invariant sequences) of the ($K(K+1)/2$)-dimensional RECIPE-feasible polytope, 
it is much more computationally efficient than HRS when $K$ is large (say $K>100$).
Luckily, many good XDD sequences still exist within this ``sub-polytope'', which can be (relatively) rapidly found by QPS.

\section{Evaluation}
\label{sec:eval}

In this section, we first compare SS (Shifted Soliton) and (codes discovered by) HRS and QPS, against PINT, the state-of-the-art 
distributed LT code.
Then, we show that RECIPE-t achieves a similar coding efficiency as RECIPE-d given a moderately large AVST (no more than 1MB).
We measure coding efficiency by the average number of codewords needed to completely decode a path (of $k$ hops).
According to our experiments, the comparison results are roughly the same when other metrics such as the $99\%$ quantile are used (see Appendix~\ref{sec:more-evaluation}). 
 


\noindent\textbf{Comparison with PINT:}
\label{sec:compare-pint}
We use the following two typical values of network diameter K from the evaluation of PINT in~\cite{ben_basat_pint_2020}: $K=36$ (US Carrier dataset), and $K=59$ (Kentucky Datalink dataset). 
The other two diameter values, $K= 118$ and $K=236$, result from fragmenting each switch ID in Kentucky Datalink to $2$ and $4$ message blocks, respectively.
As shown in \autoref{fig:compare}, SS and QPS outperform PINT consistently (i.e., for every possible path length) from 1 to 236.
HRS, however, underperforms PINT at small $k$ values when $K$ is 118 or 236 due to the (cumulative) ``inherited'' coding inefficiency (while searching backward from $K$) 
we mentioned earlier.
When the path length $k = K$, SS, HRS and QPS all outperform PINT significantly, by
24.5\% to 58.8\%, 33.4\% to 66.7\%, and 32.2\% to 63.2\%, respectively. 
QPS consistently outperforms SS, which is expected since SS is discovered by our ``naked eye'' in the same ``subpolytope'' whereas QPS is by the computer.
In the three experiments using Kentucky Datalink (in which $K\ge 59$), HRS outperforms QPS roughly when $k\ge 0.75 K$, but underperforms 
QPS at shorter path lengths
by 7.6\% (when $k = 21$ and $K=59$) to 33.5\% (when $k = 78$ and $K=236$), again due to the ``inherited'' coding inefficiency problem.

\noindent\textbf{RECIPE-t vs RECIPE-d: }\label{sec:compare-tvsd}
\autoref{fig:zoomin} shows how the (AVST) table size affects the coding efficiency with the path length $k$ varying between 1 and 59.
\autoref{fig:zoomin} contains three plots representing RECIPE-t with table sizes of 1000, 10,000, and 30,000 rows respectively and one plot representing RECIPE-d (which induces HRS).
\autoref{fig:zoomin} shows that as the table size becomes larger, the coding efficiencies of RECIPE-t becomes closer to those of RECIPE-d.
The zoomed tail section (when $k$ gets close to $K$) plotted in~\autoref{fig:zoomin} shows that the difference in coding efficiency between RECIPE-d and RECIPE-t is negligible 
when the table has 30,000 rows (about 960 KB in size).   
This is a small cost to pay (for each switch to store this table) in exchange for packets not having to carrying the XOR degree information in them.  

\section{Related Work}
\label{sec:related}
RECIPE codes are different than the so-called ``distributed LT codes''~\cite{distributedLT, distributedLT-journal} in literature. The latter codes are designed for \emph{multiple-access relay channels}, in which messages come from multiple sources that do not communicate with each other at all. These sources send LT codewords to a common relay node, which further combines (XOR-merges) them to improve the coding efficiency.
Our setting is less restrictive in the sense that sources (switches) are allowed to have some limited communication through the TTL field and the XOR degree (explicit in RECIPE-d and implicit in RECIPE-t).


\section{Conclusion}
\label{sec:conclusion}

As the first systemic study of LT codes in the distributed setting, we make the following three contributions.
First, we show that not every LT code is feasible (n the distributed setting, and
give the sufficient and necessary condition for being RECIPE-feasible.
Second, we propose RECIPE, a distributed encoding protocol that can efficiently implement any RECIPE-feasible code. 
Third, we propose two heuristic search algorithms, namely HRS and QPS, that have led to the discovery of RECIPE-feasible codes that are much more efficient than PINT, the state of the art.  

\noindent\textbf{Acknowledgment} This material is based upon work supported by the National Science Foundation under Grant No. CNS-1909048 and CNS-2007006.

	\newpage~\newpage
	\bibliographystyle{IEEEtran}
	\bibliography{IEEEabrv,reference}

\newpage\appendix
\subsection{Proof of Necessity of \autoref{main_result}}\label{pf-th31}
To prove necessity, it suffices to show by contradiction that if any triplet $(q_{i-1}(d), q_i(d), q_i(d+1))$ violates (\ref{eq:ineq}), (in other words $q_{i-1}(d) < q_i(d) + q_i(d+1)$), it is impossible to transform the XDD $q_{i-1}$ to $q_i$ using the three available actions (\Add{}, \Skip{}, or \Replace{}). 

To this end, consider a longest possible path of $K$ hops. For $i = 1,2, \cdots, K$, denote by $X_i$ the XOR set (the indices of messages contained in this packet) when it leaves switch $i$.
Let $S$ be an arbitrary subset of $[i-1] \triangleq \{1,2, \cdots, i-1\}$ that has cardinality $d$.  Then according to the the uniformity condition, we have
\begin{align*}
&	\Pr(X_{i}=S)=q_{i}(d),\\
&	\Pr(X_{i}=S\cup \{i\})=q_{i}(d+1).\\
&	\Pr(X_{i-1}=S)=q_{i-1}(d).
\end{align*}
As a result, $\Pr(X_{i-1}=S) < \Pr(X_{i}=S) + \Pr(X_{i}=S\cup \{i\})$ by our assumption above. Note that the only case in which $X_i = S$ is when $X_{i-1} = S$ and switch $i$ performs \texttt{Skip}. Also, the only case in which $X_{i}=S\cup \{i\}$ is when $X_{i-1} = S$ and switch $i$ performs \texttt{Add}. Since these two case are exclusive, and that $X_{i-1} = S$ is a necessary condition for both of them, it must be that $\Pr(X_{i-1}=S) \geq \Pr(X_{i}=S) + \Pr(X_{i}=S\cup \{i\})$, which leads to a contradiction with our assumption above.

\subsection{Proof of Correctness of APA Designation in (\ref{eq:op_prob})} \label{pf-sufficiency}
We first show all entries in~(\ref{eq:op_prob}) are well-defined, and then that they actually induce the intended XDD sequence $\vec{q}_1, \vec{q}_2, \cdots \vec{q}_K$.

The fact of being well-defined is obvious based on the following two observations. 
First, $q_{i-1}(d)$, which is the denominator of $p_A(i, d)$ and $p_S(i, d)$ in~(\ref{eq:op_prob}), is always nonzero for all meaningful APA entries, because otherwise switch $i$ will never receive a packet of XOR degree $d$ assuming the XDD $\vec{q}_{i-1}$ is induced as intended (which will we prove shortly). 
Second, as a direct corollary of the RECIPE-feasible condition~(\ref{eq:ineq}), the three APA entries $\padd{i}{d}, \pskip{i}{d}, \preplace{i}{d}$ form a well-defined discrete distribution in the sense that they are all greater than or equal to $0$, and they sum up to $1$. 


We now show inductively why this designation satisfies the uniformity condition and induces $\vec{q}_1, \vec{q}_2, \cdots \vec{q}_K$.
To begin with, the only non-empty XOR set for $i=1$ is $X_1 = \{1\}$. This is trivially induced by letting switch $1$ perform \texttt{Replace} in all cases.

Now we consider the distribution of the XOR set $X_i$ at switch $i$. Note that the induction assumption means that for $\Pr(X_{i-1} = S') = q_{i-1}(d)$ for any subset $S' \subset [i-1]$ of cardinality $d$. It suffices to show $\Pr(X_{i} = S) = q_{i}(d)$ for any subset $S \subset [i]$ of cardinality $d$.

To this end, we discuss on following three cases based the value of $S$:
\paragraph{When $i \notin S$} In this case, $X_i = S$ if and only if $X_{i-1} = S$ and switch $i$ performs \texttt{Skip}. Therefore, $\Pr(X_i = S) = \Pr(X_{i-1} = S) p_S(i, d) = q_{i-1}(d) p_S(i, d) = q_i (d)$. 
\paragraph{When $i\in S$ and $S \neq \{i\}$} In this case, $X_i = S$ if and only if $X_{i-1} = S \setminus \{i\}$ (whose cardinality is $d-1$) and switch $i$ performs \texttt{Add}. Therefore, $\Pr(X_i = S) = \Pr(X_{i-1} = \nolinebreak S\setminus\nolinebreak \{i\}) p_A(i, d-1) = q_{i-1}(d-1) p_A(i, d-1) = q_i (d)$. 
\paragraph{When $S = \{i\}$} In this case, $X_i = S$ if and only if switch $i$ performs \texttt{Replace}. Note that we do not need to prove this case given the results for the  two cases a) and b) above.
That is because the distribution of $X_i$ (which is always nonempty) is parameterized by only $2^i -2$ probabilities, since we have an addition constraint that all probabilities sum up to 1. The two cases above already covers $2^i -2$ subsets (except $\{i\}$) that is enough to show $\vec{q}_i$ is induced by the APA.

\subsection{Detailed Formulation of QPS}~\label{expl-mean-field}
In this subsection, we formulate and explain the quadratic programming (QP) procedure that QPS uses to find good XDD sequences that are both invariant and RECIPE-feasible.

Consider the following decoding process in which codewords arrive one by one at the host. Initially, all messages are unknown. The host needs to receive $T_1$ (a random variable) codewords before it can decode some massage (say the first in the decoding order). 
In the sequel, we use a different letter $j$ for decoding order to avoid confusion with the encoding order (which we denoted by the hop count $i$).
The next message ($j=2$) is decoded after $T_2$ more codewords are received, and so on. The total number of codewords needed for complete decoding (our measure for coding efficiency) is $T\triangleq T_1 + T_2 + \cdots + T_K$. 

QPS solves the following optimization problem, in which each $t_j$ (for $j=1,2,\cdots, K$)  is a ``first moment'' approximation of $E[T_j]$. 
\begin{align}
	\text{min}&\text{imize } t_1 + t_2 + \cdots + t_K, \text{where for } j=1,2,\cdots, K,\nonumber\\
	&t_j  \triangleq \left( 1 -  P^{rel}_j \cdot\sum_{\tau=0}^{i-1}t_\tau\right) / P^{suc}_j, \:  \label{eq:mean-field}\\
	&\text{$\mu$ satisfies \autoref{th:inv-point}.,} \nonumber\\
	&P^{rel}_j  \triangleq \sum_{d=2}^j (K-j+1) \binom{j-2}{d-2} \mu(d)/ \binom{K}{d}, \nonumber \\
	&P^{suc}_j  \triangleq \sum_{d=1}^j (K-j+1) \binom{j-1}{d-1}\mu(d)/ \binom{K}{d}. \nonumber
\end{align}
Problem (\ref{eq:mean-field}) can be rewritten into quadratic programming (QP), since its objective and constraints only involve summation, multiplication and division.

We now explain why (\ref{eq:mean-field}) is formulated in this way.
The distribution of $T_j$ (for $j=1,2,\cdots, K$) has two cases depending on whether the $j^{th}$ message $M_j$ can be decoded directly from the first $S_j \triangleq \sum_{\tau = 1}^{j-1} T_\tau$ codewords received by the host (that are used to decode the first $j-1$ messages). If yes, then obviously $T_j = 0$; otherwise $T_j$ is a geometric random variable with success probability $P^{suc}_j$.
In the rest of this subsection, we provide more details to each of the two cases.

\subsubsection{$M_j$ can be decoded without needing more codewords} This case happens only when the following three conditions are satisfied. First, least one codeword containing $M_j$ is received among the first $S_j$ codewords. Second, $M_j$ in such codeword(s) above is blocked by one or more messages upon the reception. (Otherwise, $M_j$ would have been decoded earlier.) Third, the last blocking message is exactly $M_{j-1}$, whose decoding \emph{releases} $M_j$ from such codeword(s) above.

Supposing $M_j$ is contained in a codeword $cw$ with XOR degree $d$, then the probability that $M_j$ is released by the decoding of $M_{j-1}$ (but not by any previous message) can be computed using the following ball-shuffling model.
Consider $K$ balls (messages) placed in a linear order such that the first $j-1$  balls correspond to the first $j-1$ messages in the order of decoding. 
$M_j$ is in the  remaining $K-j+1$ balls, but it is unknown at the decoding of $M_{j-1}$ which one is $M_j$.
The codeword $cw$ contains a random subset of cardinality $d$ out of these $K$ balls. 
Equivalently, we randomly color $d$ balls in red (that are contained in $cw$) and leave the other $K-d$ balls uncolored.
Then, the release probability is equal to the probability that exactly $d-2$ red balls are among the first $j-2$ balls, the $(j-1)^{th}$ ball is red (so that the decoding of $M_{j-1}$ releases $M_j$), and the remaining red ball is in the last $K-j+1$ balls (that red ball is $M_j$). 
This probability, which we denote by $p^{rel}_j(d)$, is equal to $(K-j+1) \binom{j-2}{d-2}/ \binom{K}{d}$ by basic combinatorics. Then, the overall release probability when the XOR degree $d$ is drawn from the XDD $\vec{\mu}$ is 
$P^{rel}_j = \sum_{d=2}^j p^{rel}_j(d)\mu(d)$.

In (\ref{eq:mean-field}), we approximate the probability of Case 1) (which we denote here by $p^f_j$) by $S_j P^{rel}_j$, i.e., the expected number of releases among the $S_j$ received codewords at the decode of $M_{j-1}$. 
The reason here is not obvious, because it is possible that the decoding of $M_{j-1}$ releases two (or more) messages.
We first consider ``double release'' events. 
Regarding such ``double release'', the exact (correct) way of counting $p^f_j$ is to add the double release probability of $M_{j-2}$ and that of $M_{j-1}$ (both of which can release $M_j$ without needing more codewords).
To make life easier, we assume the decoding is a ``quasi-stationary'' process (changing slowly with $j$) in the sense that the two double release probabilities above are approximately equal.
As a result, we add twice the double release probability of  $M_{j-1}$ to the counter for $p^f_j$.
Similarly, for 	``multiple release'' events, we add its probability times the number of releases to the counter for $p^f_j$.
Moreover, we always assume multiple releases, when they happen, never collide on the same message, so that they are always counted independently to $p^f_j$.
This approximate way of counting $p^f_j$ is obviously the same as the way of counting the expected number of releases at the decoding of $M_{j-1}$. This is why we approximate the former value by the latter.

\noindent\textbf{Remark}: The latter assumption above (no collision) can be mitigated by adding a second order term to the estimation of $p^f_j$ (that compensates for possible collisions) so that it becomes $S_j P^{rel}_j - 1/2 \cdot (S_j P^{rel}_j )^2 / (K - j +1)$. 
In our experiments, such a correction only increases the coding efficiency of QPS by about 5\%, but more than doubles the search time (because the formulated QP optimization has more variables). This is why we suggest using only  the first-order approximation above.

\subsubsection{More codewords are needed to decode $M_j$}
 In this case, the only way to decode $M_j$ is to receive a new codeword that contains $M_j$ as the only unknown message. Since each received codeword is independent, the number of required codewords follows geometric distribution. $P^{suc}_j$, the success probability of receiving a codeword that satisfies the condition above can be similarly computed using a similar ball-shuffling model.
 The probability $p^{suc}_j(d)$, of success upon receiving a codeword of XOR degree $d$, is equal to the probability that given $K$ randomly shuffled balls, in which $d$ are red (contained in codeword), the first $j-1$ balls (the first $j-1$ decoded messages) contain exactly $d-1$ red balls, and (of course) the last red ball is among the last $K-j+1$ balls. This probability is $(K-j+1) \binom{j-1}{d-1}/ \binom{K}{d}$. Then, the overall success probability for a random codeword whose degree $d$ is sampled from XDD $\vec{\mu}$ is 
 $P^{suc}_j = \sum_{d=1}^j p^{suc}_j(d)\mu(d)$.

\subsection{More Examples and Results}\label{sec:more-evaluation}
\begin{figure}[!htb]
	\centering
	\includegraphics[width=0.6\linewidth]{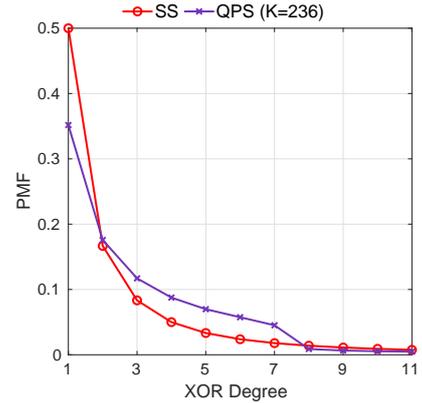}
	\caption{Comparison of PMF on XOR degrees between 1 and 11.}
	\label{fig:xdd-compare}
\end{figure}
\begin{figure*}[!htb]
	\centering
	\includegraphics[width=\linewidth]{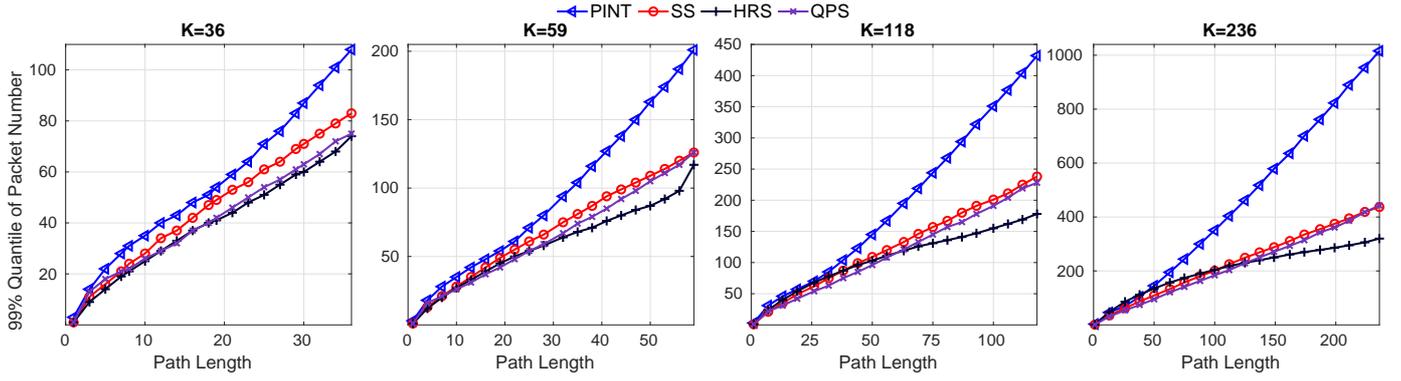}
	\caption{Comparison of coding efficiency in terms of $99\%$ quantile.}
	\label{fig:99quantile}
\end{figure*}
\noindent\textbf{A ``good'' XDD found by QPS:} 
\autoref{fig:xdd-compare} shows the PMF of a ``good'' RECIPE code found by QPS compared against SS (Shifted Soliton), the one we discovered accidentally. This  QPS code is optimized for the network diameter $K=236$ while QPS returns similar codes in all other diameters used in our evaluation. We only show the probability mass on the first 11 XOR degrees, since the remaining values ($d$ from 12 to 256) are very similar for both codes.

Compared with SS, the QPS code has much less mass on $d=1$ (0.352 vs 0.5), but has much more mass when $3\le d \le 7$. It is also obvious that QPS attempts to achieve a much more gentle PMF curve than SS in the sense that the RECIPE-feasible condition defined in \autoref{th:inv-point} is an equation (without slack) for the QPS code when $d\le 7$.
As a result, we are convinced that the most crucial part of a ``good'' RECIPE code is the XDD on the first few XOR degrees. QPS seems to have reached a near-optimal tradeoff between the Soliton code (which has a spike at $d=2$) and the RECIPE-feasible condition that disallows spikes in invariant RECIPE codes (see~\autoref{th:inv-point}).
When it comes to RECIPE codes outside the sub-polytope of invariant codes, we know HRS codes can have spikes (being Robust Soliton), which means better coding efficiency than invariant codes, at the start of search. However, it is still a challenge to find out codes (even if they exist) that still has such a good PMF shape in later iterations since the errors are inherited.

\noindent\textbf{99\% Quantile Coding Efficiency:} 
\autoref{fig:99quantile} compares the coding efficiencies of different XDD sequences in terms of the $99\%$ quantile of the packets needed to decode the path information. It shows almost the same trends as \autoref{fig:compare}. The only major difference here is that the outperformance of HRS over the other two RECIPE codes (SS and QPS) when the path length is close to $K$ is more significant.
This is because the search process of HRS starts from Robust Soliton, which is known to be able to decode without too many codewords even in the worst case. 
In contrast, the other two codes are designed mainly to optimize the average performance, and it is still an open problem how to adjust their designs for better worst-case performance.	
\end{NoHyper}

\end{document}